\documentclass[aps, 12pt, final, notitlepage, oneside, onecolumn, nobibnotes,
nofootinbib,superscriptaddress, noshowpacs,amsmath,amssymb]
{revtex4}
\usepackage{bm}
\usepackage{epsfig}
\usepackage{graphics}

\usepackage[english]{babel}
\begin{document}
	
\title{QUASI-SOLID STATE MICROSCOPIC DYNAMICS \\ IN EQUILIBRIUM CLASSICAL LIQUIDS: \\ SELF-CONSISTENT RELAXATION THEORY}
	
\author{ \firstname{A.~V.}~\surname{Mokshin}}
\affiliation{Kazan Federal University, 420008 Kazan, Russia}
	
\author{\firstname{R.~M.}~\surname{Khusnutdinoff}}
\affiliation{Kazan Federal University, 420008 Kazan, Russia}
	
\author{\firstname{Ya.~Z.}~\surname{Vilf}}
\affiliation{Kazan Federal University, 420008 Kazan, Russia}
	
\author{\firstname{B.~N.}~\surname{Galimzyanov}}
\affiliation{Kazan Federal University, 420008 Kazan, Russia}
	
%\pacs{61.20.-p,61.05.F-,67.25.du}
\today
	
\begin{abstract}
In the framework of the concept of time correlation functions, we develop a self-consistent relaxation theory of the transverse collective particle dynamics in liquids. The theory agrees with well-known results in both the short-wave (free particle dynamics) and the long-wave (hydrodynamic) limits. We obtain a general expression for the spectral density~$C_T(k,\omega)$  of transverse particle current realized in the range of wave numbers $k$. In domain of microscopic spatial scales comparable to action scale of effective forces of interparticle interaction, the theory reproduces a transition from a regime with typical equilibrium liquid dynamics to a regime with collective particle dynamics where properties similar to solid-state properties appear: effective shear stiffness and transverse (shear) acoustic waves. In the framework of the corresponding approximations, we obtain expressions for the spectral density of transverse particle current for all characteristic regimes in equilibrium collective dynamics. We obtain expressions for dispersion law for transverse (shear) acoustic waves and also relations for the kinematic shear viscosity $\nu$, the transverse speed of sound $v^{(T)}$, and the corresponding sound damping coefficient $\Gamma^{(T)}$. We compare the theoretical results with the results of atomic dynamics simulations of liquid lithium near the melting point.		
\end{abstract}
\maketitle

\section{Introduction}
The absence of shear stiffness is one of key physical-mechanical properties of equilibrium classical liquids distinguishing liquids from solids~\cite{Frenkel_book,Bol_Phys_Enciclopedy,Phys_Enc}. This property is manifested in the inability of liquids to maintain a definite volume form if there are no boundary surfaces and the ability to take the shape of a filled vessel. The absence of shear stiffness is also manifested in the fact that in contrast to crystalline and amorphous solids, liquids are unable to support the propagation of transverse (so-called shear) oscillation processes~\cite{Trachenko2016,Tareyeva2018}. In fact, everything occurs for equilibrium classical liquids on that way only under conditions close to hydrodynamic, when we consider macroscopic spatial scales and sufficiently large observation times in corresponding experiments. Thus, for example, in the case of a high-frequency shear-deformation perturbation under the action of a varying force, a high-density liquid demonstrates an elastic response typical for solids. This is clearly seen in experiments measuring the frequency-dependent shear modulus~\cite{hypersound_experiments}. Such a reaction of a liquid to a perturbation is fully rather understandable: in simple liquids, this reaction is most clear expressed when external mechanical excitation has a time scale $\tau'$ (where $\tau' = 2\pi/\omega'$ and $\omega'$ is the excitation frequency) comparable to the structural relaxation time $\tau_s$. Moreover, transverse oscillatory processes typical for solids can be manifested in the collective dynamics of particles of a high-density liquid at \textit{microscopic spatial scales}. In this case, the manifestation of microscopic shear stiffness is due to the action of \textit{effective attractive forces} in the interaction between particles of liquid, and the spatial scale $l_g$ at which liquid quasistiffness is manifested must be determined by the size of the action region of these forces.
	
The existence of propagating shear waves in compressed noble gases (so-called Lennard-Jones liquids in the general case) near the triple point was first confirmed using methods for modeling equilibrium molecular dynamics~\cite{Levesque/Verlet_1973,Sjogren1978}. The spectra of the computed spectral density $C_T(k,\omega)$ of transverse molecule current for wave numbers $k$ corresponding to spatial scales of several effective linear sizes of the molecule turned out to contain in expressed component in the terahertz region with a maximum at the frequency $\omega_c^{(T)} \neq 0$~\cite{Levesque/Verlet_1973}. The dependence of $\omega_c^{(T)}$ on the wave number $k$ produces a dispersion relation $\omega_c^{(T)}(k)$ characterizing the propagation of transverse oscillatory processes in a liquid. As a rule, there is a \textit{gap} with zero values of $\omega_c^{(T)}$ in the dispersion law $\omega_c^{(T)}(k)$ in the domain of small wave numbers $k$. Obviously, the size $k_g$ of this gap must be related to the spatial scale $l_g$ at which a quasistiff shear response is manifested in a liquid.
	
The methods of classical and quantum mechanical molecular dynamics simulations still remain the most appropriate tool for directly quantitatively estimating almost all effects related to the shear oscillation dynamics of molecules in both model~\cite{Donko,Khrapak2019,Ryltsev2013} and real liquids (see, e.g.,~\cite{Gonzalez2017,Jakse/Bryk2019,Jones2016,Fomin2019,Wang2019}). Although direct experimental measurements are difficult, some indirect information about shear waves in high-density liquids can be obtained from inelastic neutron scattering (INS) and inelastic x-ray scattering (IXS) experiments~\cite{Hosokawa_2015}. In particular, effects due to the transverse oscillation dynamics of particles can be manifested in the contour of the dynamic structure factor $S(k,\omega)$ measured in INS and IXS experiments~\cite{Egelstaff_book,Burkel_review}.
	
We note that key components (a central Rayleigh and two symmetric Brillouin components) of the scattering law in a liquid at microscopic spatial scales accessible by INS and IXS experimental techniques are not separated, and the contribution to the general scattering law $S(k,\omega)$ due to transverse oscillation dynamics is weakly expressed and is not recognized in the form of a separate component. Features typical for transverse oscillatory processes must obviously appear in the spectra $S(k,\omega)$ in a frequency range between the Rayleigh and Brillouin peaks. In fact, only continuations of the Brillouin and Rayleigh peaks are observed in the experimental spectra $S(k,\omega)$ in this frequency range. Consequently, directly interpreting the experimental data is complicated, and the resulting conclusions from this interpretation can be ambiguous~\cite{Hosokawa_Ga,Hosokawa2013,Monaco_PNAS,Monaco_PRB}.
	
We mention some currently well-known theoretical models and approaches connected with describing transverse collective dynamics: the viscoelasticity model~\cite{MacPhail/Kivelson,Bryk_PRE2000}, the generalized collective mode method~\cite{Mryglod_CMP1994}, the Maxwell-Frenkel model for describing the dispersion relation of transverse collective modes~\cite{Trachenko2017,Baggioli2020}, and the approach for recovering the dispersion dependences of longitudinal and transverse collective modes based on separate and joint mode analysis (the ``two oscillator model'')~\cite{Yurchenko2019,Yurchenko_SC_2019,Yurchenko_1}. It is also remarkable that the presence of transverse excitations can lead to the appearance of so-called positive dispersion of longitudinal sound: the observed/measured speed of sound in a certain range of finite wave numbers is higher than the hydrodynamic values~\cite{Fomin_JPCM_2016,Brazhkin_2018}.
	
Here, we show that the transverse collective dynamics in simple liquids can be described in the framework of a self-consistent relaxation theory. This theoretical description is a direct development of the idea of the time-scale invariance of relaxation processes in liquids~\cite{Mokshin_PRE2001,Mokshin2002}, based on which a theory of local density fluctuations in liquids was developed and the use of this theory for analyzing INS and IXS experimental data was demonstrated in liquid lithium~\cite{Mokshin_JPCM2018}, sodium~\cite{Mokshin_JPCM2003,Mokshin_JCP2004}, cesium~\cite{Mokshin_PRE2001,Mokshin2002}, aluminum~\cite{Mokshin_JETP2006, Mokshin_JPCM2007}, and gallium~\cite{Khusnutdinoff2020}. We present the theoretical formalism in Sec.~\ref{sec: theory} and compare the theoretical results with data from equilibrium atomic dynamics simulations of liquid lithium in Sec.~\ref{sec: comparison}.

\section{Theoretical formalism \label{sec: theory}}
In the framework of the general statistical mechanics approach, we can regard an equilibrium liquid as a multiparticle system, an ensemble of atoms and molecules~\cite{Ryzhov_UFN2008}. In this case, it is convenient to use the mathematical apparatus of correlation functions, distribution functions, moments, and cumulants of these functions to describe both structural and dynamical properties of this liquid~\cite{Hansen/McDonald,Zwanzig2001,Mokshin/Yulmetyev,Klumov}.
	
We consider an isotropic system comprising $N$ identical classical particles of mass $m$ in a volume $V$. The coordinates $\mathbf{r}_j$ and velocities $\mathbf{v}_l$, $j,\,l = 1,\, \ldots,\, N$, of these particles determine a $6N$-dimensional phase space, where we can define a certain dynamical variable $A(\mathbf{r}_1,\, \ldots,\, \mathbf{r}_N,\, \mathbf{v}_1,\, \ldots,\, \mathbf{v}_N)$. Let the system Hamiltonian $H$ be given. We can then determine the mean $\left \langle A \right \rangle$ in terms of the phase space density distribution $\rho \propto \exp\{-(H-\mu N)/(k_BT)\}$, where $T$ and $\mu$ are the temperature and chemical potential. Moreover, the Hamiltonian $H$ determines the evolution of the dynamical variable $A$:
\begin{equation}\label{43}
\frac{\partial}{\partial t} A(t) = \left \{ H, A(t) \right \} = i\hat{\mathcal{L}}A(t),
\end{equation}
where $\{...\}$ is Poisson bracket, $\hat{\mathcal{L}}$ is the Liouville operator, which is Hermitian~\cite{Zwanzig2001,Balucani2003}. In the case where the interaction potential $u(r)$ between particles is pairwise and spherical, we can write the Liouville operator in the form
\begin{equation}\label{5}
\hat{\mathcal{L}} = -i\left\{ \frac{1}{m}\sum_{j=1}^{N} \left ( \mathbf{p}_j\cdot\nabla_j \right ) - \sum_{l>j=1}^{N}\nabla_ju(r_{jl}) \left ( \nabla_{\mathbf{p}_{j}}-\nabla_{\mathbf{p}_{l}} \right ) \right\}.
\end{equation}
where $\mathbf{p}_{j}$ is the momentum of the $j$th particle and $\nabla_j$ and $\nabla_{\mathbf{p}_{j}}$ are the coordinate and momentum gradients.

\subsection{Dynamical variables \label{sec: Dyn_var}}
Let the particle current occur in some region of the inverse space of the wave vectors $\textbf{k}$ and be determined by the velocities of all particles belonging to that region. We can write the expression for the full particle current
\begin{eqnarray}\label{44}
\textbf{j}_{k}(t) = \sum_{l=1}^{N} \textbf{v}_{l}(t)  \; \mathrm{e}^{-i \textbf{k}\cdot \textbf{r}_{l}(t)},
\end{eqnarray}
where $k = \left | \textbf{k} \right |$ is the wave number and $\textbf{r}_{l}(t)$ is the radius vector characterizing the trajectory of the $l$th particle. Assuming that the wave vector $\textbf{k}$ is directed along the coordinate axis $OZ$, we can introduce the corresponding expressions for the longitudinal current
\begin{eqnarray}\label{45}
j_{k}^{(z)}(t) = \sum_{l=1}^{N} v_{l}^{(z)}(t)\; \mathrm{e}^{-i\textbf{k}\cdot \textbf{r}_{l}(t)}
\end{eqnarray}
and the transverse current
\begin{eqnarray}\label{46}
j_{k}^{(x)}(t) = \sum_{l=1}^{N} v_{l}^{(x)}(t)\; \mathrm{e}^{-i\textbf{k}\cdot \textbf{r}_{l}(t)},
\end{eqnarray}
and
\begin{eqnarray}\label{46}
j_{k}^{(y)}(t) = \sum_{l=1}^{N} v_{l}^{(y)}(t)\; \mathrm{e}^{-i\textbf{k}\cdot \textbf{r}_{l}(t)},
\end{eqnarray}
where $v^{(x)}$, $v^{(y)}$ and $v^{(z)}$ are the velocity components for the $l$th particle, $\mathbf{v}_l = v_l^{(x)}\mathbf{i_1} + v_l^{(y)}\mathbf{i_2} + v_l^{(z)}\mathbf{i_3}$ and $\textbf{j}_{k} = j_{k}^{(x)}\mathbf{i_1} +  j_{k}^{(y)}\mathbf{i_2}   + j_{k}^{(z)}\mathbf{i_3} $. The quantities $j_{k}^{(z)}$, $j_{k}^{(x)}$ and $j_{k}^{(y)}$ and similarly for $\textbf{j}_{k}$ have the meaning of dynamical variables characterizing the collective dynamics of the multiparticle system. The transverse current components $j_{k}^{(x)}$ and $j_{k}^{(y)}$ in a structurally isotropic system (dense gases, liquids, amorphous solids) are obviously equal.
	
It is convenient in considering transverse collective dynamics to choose the transverse current $A_1^T$ as initial dynamical variable:
\begin{equation}
A_1^T(k) = j_{k}^{(x)}= j_{k}^{(y)}.
\label{eq: initial_dyn_var}
\end{equation}
Using the orthogonalization procedure~\cite{Reed_Book}
\begin{eqnarray} \label{eq: orthogonal}
A_{l+1}^T(k) &=& i\hat{\mathcal{L}}A_l^T(k)+ \frac{\left \langle \left | A_{l+1}^T(k) \right |^2 \right \rangle}{\left \langle \left | A_{l}^T(k) \right |^2 \right \rangle} A_{l-1}^T(k), \\
l &=& 1,\; 2,\; 3,...;  \ \  A_{0}^T \equiv 0, \nonumber
\end{eqnarray}    	
we obtain the set
\begin{equation} \label{eq: set_var}
\textbf{A}^T(k) = \left\{A_1^T(k),\; A_2^T(k),\; A_3^T(k),\; \ldots,\; A_l^T(k),\; \ldots \right\},
\end{equation}    	
of dynamical variables satisfying the orthonormalization condition
\begin{eqnarray} \label{eq: orthonorm}
\left \langle A_{i}^T(-k)A_j^T(k) \right \rangle &=& \delta_{i,j}\left \langle \left | A_j^T(k) \right |^2 \right \rangle, \\
i,\;j &=& 1,\; 2,\; 3,\; \ldots. \nonumber
\end{eqnarray}
where the angle brackets denote averaging over the statistical ensemble and $\delta_{i,j}$ is the Kronecker symbol.
	
It follows from expressions~(\ref{eq: initial_dyn_var}) and (\ref{eq: orthogonal}) that the dynamical variable $A_2^T(k)$ is related to the acceleration
and kinetic energy in the transverse collective particle dynamics:
\begin{equation}
\label{eq: A2}
A_2^T(k) =  \sum_{l=1}^{N} \dot{v}_l^{(x)}\; \mathrm{e}^{-i\mathbf{k}\cdot\mathbf{r}_l} - i \sum_{l=1}^{N} k_x \left (v_l^{(x)}\right )^2\; \mathrm{e}^{-i\mathbf{k}\cdot\mathbf{r}_l},
\end{equation}
For the third dynamical variable, we obtain
\begin{eqnarray}
\label{eq: A3}
A_3^T(k) &=& \sum_{l=1}^{N} \ddot{v}_l^{(x)}\; \mathrm{e}^{-i\mathbf{k}\cdot\mathbf{r}_l} - 3i \sum_{l=1}^{N} k_x \dot{v}_l^{(x)} v_l^{(x)}\; \mathrm{e}^{-i\mathbf{k}\mathbf{r}_l} \nonumber \\
&-& \sum_{l=1}^{N} k_x^2 \left ( v_l^{(x)}\right )^3\; \mathrm{e}^{-i\mathbf{k}\cdot\mathbf{r}_l}
+ \frac{\left \langle \left | A_{2}^T(k) \right |^2 \right \rangle}{\left \langle \left | A_{1}^T(k) \right |^2 \right \rangle} \; \sum_{l=1}^{N} {v}_l^{(x)}\; \mathrm{e}^{-i\mathbf{k}\cdot \mathbf{r}_l}.
\end{eqnarray}
The dynamic variable $A_3^T(k)$ characterizes the change in the particle acceleration.

\subsection{Correlation functions, frequency moments, and relaxation parameters}
We define time correlation functions (TCFs) for the set $\textbf{A}^T(k)$ of dynamical variables:
\begin{equation}
\label{eq: TCF}
M_{l}^T(k, t) = \frac{\left \langle A_{l}^T (-k,0) A_{l}^T(k, t) \right \rangle}{\left \langle \left | A_{l}^T(k, 0) \right |^2 \right \rangle}, \ \ \ l = 1,\; 2,\; \ldots ,
\end{equation}
where
\[
M_1^T(k,t) \equiv C_T(k,t).
\]
The TCFs $M_{l}^T(k, t)$ have the properties~\cite{Plakida2005,Lee2000}
\begin{subequations}\label{tq: TCF_prop}
\begin{equation}
M_{l}^T(k,t) \bigg|_{t=0} = 1, \label{eq: prop1}
\end{equation}
\begin{equation}
\left\{\begin{matrix} \text{ }\cfrac{d^{p}}{dt^p}M_{l}^T(k,t)\bigg|_{t=0} = 0, \text{$p$ is odd,}\\
\cfrac{d^{p}}{dt^p}M_{l}^T(k,t)\bigg|_{t=0} \neq 0, \text{$p$ is even,}
\end{matrix}\right. \label{eq: prop2}
\end{equation}
\begin{equation}
\left | M_{l}^T(k,t) \right |\leq 1, \label{eq: prop3}
\end{equation}
\begin{equation}
\lim_{t \rightarrow \infty} M_{l}^T(k, t) = 0. \label{eq: prop4}
\end{equation}
\end{subequations}
Properties~(\ref{eq: prop1}), (\ref{eq: prop2}) and (\ref{eq: prop3}) reveal that the TCF $M_{l}^T(k, t)$ takes its maximum value $1$ at the initial instant. Property~(\ref{eq: prop4}) is related to ergodicity and indicates that the correlation weakens with time.
	
The TCF of the transverse current is then
\begin{eqnarray}
\label{eq: CTCF}
M_1^T(k,t) = C_T(k,t) = \frac{\left \langle j_{-k}^{(x)}(0) j_{k}^{(x)}(t) \right \rangle}{\left \langle \left |j_{k}^{(x)}(0) \right | ^2 \right \rangle}.
\end{eqnarray}
This quantity is related to the spectral density of transverse current TCF by
\begin{eqnarray}
\label{eq: spectr_density_CTCF}
C_T(k,\omega) = \frac{1}{2\pi}\int_{-\infty}^{\infty} \; C_T(k,t)\; \mathrm{e}^{i\omega t}\;dt.
\end{eqnarray}
The quantity $\left \langle \left |j_{k}^{(x)}(0) \right | ^2 \right \rangle$ in~(\ref{eq: CTCF}) is equal to
\begin{equation}
\left \langle \left |j_{k}^{(x)}(0) \right | ^2 \right \rangle = \omega_T^{(0)}(k) =  \frac{1}{N} \sum_{l=1}^N \left (v_l^{(x)} \right )^2 = \frac{k_BT}{m}.
\end{equation}
	
We define the frequency moments of the spectral density $C_T(k,\omega)$ as
\begin{eqnarray}
\label{eq: freq_moments}
\left \langle \omega_T^{(l)}(k)  \right \rangle = \left ( -i \right )^l \left . \frac{d^l}{dt^l} C_T(k,t)\right |_{t=0} = \frac{\int_{-\infty}^{\infty}\omega ^l C_T(k, \omega)\;d\omega}{\int_{-\infty}^{\infty} C_T(k, \omega)\;d\omega}, \ \ \ l = 1,\;2,\;\ldots.
\end{eqnarray}
As follows from property~(\ref{eq: prop2}), only even frequency moments $\omega_T^{(2)}(k)$, $\omega_T^{(4)}(k)$, $\omega_T^{(6)}(k)$, $\ldots$ take nonzero values.
	
We introduce the frequency relaxation parameters
\begin{eqnarray}
\label{eq: freq_parameters}
\Delta_{T,\;l+1}^{2}(k) =  \frac{\left \langle \left | A_{l+1}^T(k) \right |^2 \right \rangle}{\left \langle \left | A_{l}^T(k) \right |^2 \right \rangle}, \ \ \ l=1,\;2,\;3,\;\ldots.
\end{eqnarray}    	
The super- and subscripts $T$ mean that parameters are computed based on the dynamical variables related to the transverse dynamics, and the index $l$ is the order number of the parameter according to set $\mathbf{A}^T(k)$ of dynamical variables.
	
The parameters $\Delta_{T,\;l}^2(k)$ have the following properties~\cite{Mokshin_JPCM2018}.\footnote{The parameters $\Delta_{T,\;l}^2(k)$ are called static correlation functions in the Mori formalism~\cite{Balucani2003} and denote recursion coefficients in the M.H. Lee's recurrent relation method~\cite{Lee2000}.} First, the $l$th-order frequency relaxation parameter characterizes the relaxation time $\tau_{T,\;l}(k)$ of a process related to the dynamical variable $A_l^T(k)$:
\begin{equation}
\label{eq: relaxation_time}
\tau_{T,\;l}(k)=1/\sqrt{\Delta_{T,\;l}^2(k)}.
\end{equation}
Hence, for example, as follows from~(\ref{eq: initial_dyn_var}), (\ref{eq: A2}) and (\ref{eq: freq_parameters}), the frequency relaxation parameter $\Delta_{T,\;2}^2(k)$ in the short-wave limit related to free particle dynamics characterizes mean travel time of a particle at the spatial scale $\ell=2\pi/k$:
\begin{equation}
\tau_{T,\;2}(k) = 1/\sqrt{k_x^2 \; (k_BT)/m} = 1/\sqrt{(k_x \bar{v}_x)^2},
\end{equation}
where $k_x \equiv k$ and $\bar{v}_x$ is the mean velocity of the particle. This property was discussed in detail in paper~\cite{Mokshin2005}.
	
Secondly, the frequency relaxation parameters are related to the frequency moments~\cite{Mokshin2015}:
\begin{subequations}
\label{eq: Delta_vs_moments}
\begin{equation}\label{eq: Delta1}
\Delta_{T,\;2}^2(k) = \langle \omega_T^{(2)}(k) \rangle,
\end{equation}
\begin{equation}\label{eq: Delta2}
\Delta_{T,\;3}^2(k) = \frac{\langle \omega_T^{(4)}(k) \rangle}{\langle \omega_T^{(2)}(k) \rangle} - \langle \omega_T^{(2)}(k) \rangle,
\end{equation}
\begin{equation}\label{eq: Delta3}
\Delta_{T,\;4}^2(k) = \frac{\langle \omega_T^{(6)}(k) \rangle \langle \omega_T^{(2)}(k) \rangle - (\langle \omega_T^{(4)}(k) \rangle)^2}{\langle \omega_T^{(4)}(k) \rangle \langle \omega_T^{(2)}(k) \rangle - (\langle \omega_T^{(2)}(k) \rangle)^3}.
\end{equation}
\end{subequations}
\noindent
Expressions~(\ref{eq: Delta_vs_moments}) are also called the \textit{sum rules} of the spectrum~$C_T(k,\omega)$. The mathematical meaning of this property is that a unique set of frequency relaxation parameter values can be related to a particular form of the spectrum~$C_T(k,\omega)$. Consequently, the sum rules can be used to verify the correctness of a proposed theoretical model of the spectrum~$C_T(k,\omega)$. In this case, the frequency relaxation parameter values obtained based on some theoretical model are related to the parameter values obtained from the definition of the frequency moments by numerically computing the integral expressions for the frequency moments based on an experimental spectrum~$C_T(k,\omega)$ [see the second equality in (\ref{eq: freq_moments})].
	
Finally, as follows from~(\ref{eq: initial_dyn_var}), (\ref{eq: A2}) and (\ref{eq: A3}), the frequency relaxation parameters can be expressed in terms of microscopic characteristics of the system, such as the mean particle velocity, the interaction potential between particles, and particle distribution functions~\cite{Balucani2003}. Hence, for example, in the case of the set of dynamical variables defined by~(\ref{eq: initial_dyn_var}), (\ref{eq: A2}) and (\ref{eq: A3}), the frequency relaxation parameter~$\Delta_{T,\;2}^2(k)$ is
\begin{equation}
\label{eq: Delta1_micro}
\Delta_{T,\;2}^2(k) = \left ( \frac{k_B T}{m} \right  )^2 k^2 + \frac{\rho}{m} \int \left [ 1 - \cos(kz) \right ] \frac{\partial^2 u(r) }{\partial x^2} g(r)\; d^3 \mathbf{r},
\end{equation}
where $\rho$ is the quantitative particle density and $g(r)$ is the radial distribution function of particles in the system. We can also obtain microscopic expressions for the other frequency relaxation parameters $\Delta_{T,\;3}^2(k)$, $\Delta_{T,\;4}^2(k)$, $\ldots$~\cite{Balucani2003}. These microscopic expressions are usually not used for direct numerical calculations, because they contain multiparticle distribution functions and the well-known problem of decoupling an equation chain for the particle distribution functions must be solved for corresponding computations~\cite{Bogolyubov_book}. We note that these frequency relaxation parameters can also be computed based on basic definition~(\ref{eq: freq_parameters}) with configuration data (trajectories and particle velocities) obtained, for example, from molecular dynamic modeling (see the appendix in~\cite{Mokshin_JPCM2018}). The following analysis of the frequency relaxation parameters values allows defining such physical characteristic of the system as the sound propagation velocity, the sound damping coefficient, the mean particle velocity, etc.~\cite{Mokshin_JPCM2018}.

\subsection{The function $C_T(k,t)$ in the hydrodynamic limit}
In the case of transverse collective dynamics in a liquid, solutions are generally known in only two limit cases: in the hydrodynamic limit ($k \to 0$) and in the free-moving particle limit (large wave numbers $k$).
	
In the hydrodynamic limit, the time evolution of the transverse current $j_k^{(x)}$ is given by
\begin{equation}
\label{eq: hydro1}
\frac{\partial }{\partial t} j_k^{(x)}(t) = - \nu k^2 j_k^{(x)}(t),
\end{equation}
where $\nu$ is kinematic shear viscosity. The solution of this equation has the form
\begin{eqnarray}
\label{eq: hydro_sol}
j_k^{(x)}(t) = j_k^{(x)}(0)\; \exp{(- \nu k^2 t)}.
\end{eqnarray}
Multiplying~(\ref{eq: hydro1}) by the complex conjugate quantity
\[
j_{-k}^{(x)}(0)\frac{1}{\left \langle \left |j_{k}^{(x)}(0) \right | ^2 \right \rangle}
\]
and averaging over the ensemble, we obtain an equation for the TCF $C_T(k, t)$
\begin{equation}
\label{eq: hydro2}
\frac{\partial }{\partial t}C_T(k,t) = - \nu k^2 C_T(k,t).
\end{equation}
The solution of~(\ref{eq: hydro2}) has the form
\begin{eqnarray}\label{62}
C_T(k,t) = \exp{(-\nu k^2t)}.
\end{eqnarray}
Acting on~(\ref{62}) with the Fourier transform operator
\[
FT\{f\}(\omega) = f(\omega) = \frac{1}{2\pi} \int_{-\infty}^{\infty} \exp(i\omega t) f(t)\, dt,
\]
we obtain an expression for spectral density of the TCF of the transverse current:
\begin{equation}
\label{eq: hydro3}
{C}_T(k, \omega) =  \frac{1}{\pi} \frac{\nu k^2}{\omega^2 + \left (\nu k^2  \right )^2}.
\end{equation}
Therefore, in the hydrodynamic limit, the TCF $C_T(k,t)$ is characterized by an exponentially decaying time-dependence with the relaxation time $\tau=1/(\nu k^2)$ and a Gaussian dependence on $k$. The absence of high-frequency components in the spectrum ${C}_T(k, \omega)$ defined by~(\ref{eq: hydro3}) indicates the absence of transverse acoustic waves in the hydrodynamic mode.
	
\subsection{The function $C_T(k,t)$ in the short-wave limit \label{sec: shortwavelengths}}
We consider the short-wave limit, where the particle dynamics occurs on spatial scales comparable to the mean free path. In this case, the time and spatial scales are so small that particles in fact demonstrate free motion. In the case of free motion of an arbitrary $i$th particle, its velocity is constant:
\[
v_{i}^{(x)}(t) = v_{i}^{(x)}.
\]
For the TCF of the transverse current, we have~\cite{Hansen/McDonald}
\begin{eqnarray}
\label{eq: free_part_limit}
C_T(k,t) &=& \frac{m}{k_B T} \frac{1}{N} \sum_{i,j=1}^{N} \; \left ( v_{i}^{(x)} \right )^2 \; \exp \left ( -i \textbf{k} \cdot \left [ \textbf{r}_j(t) - \textbf{r}_i(0)  \right ] \right ) \nonumber\\
&=&  \exp{\left (- \frac{k_BT}{2m} \; k^2t^2 \right ) }.
\end{eqnarray}
For the spectral density ${C}_T(k,\omega)$, we find
\begin{eqnarray}
\label{eq: spectral_density_CT}
{C}_T(k,\omega) &=&  \sqrt{ \frac{m}{2\pi \; k_BT \; k^2} }\; \exp{\left ( - \frac{m\; \omega^2}{2k_BTk^2} \right ) } \nonumber\\
&=& \sqrt{\frac{1}{2\pi}} \frac{1}{\bar{v}_x k} \; \exp{\left ( - \frac{\omega^2}{2  \bar{v}_x^2 k^2} \right )} .
\end{eqnarray}
Similarly to~(\ref{eq: hydro3}), relation~(\ref{eq: spectral_density_CT}) defines a spectrum with no high-frequency components.

\subsection{Transverse current in a range of wave numbers}
From the Liouville equations of motion for the dynamical variables $A_1^T(k)$, $A_2^T(k)$, $A_3^T(k)$ and $A_4^T(k)$ [see expressions (\ref{eq: initial_dyn_var}), (\ref{eq: A2}) and (\ref{eq: A3})] we can obtain exactly two interrelated kinetic integro-differential equations containing the TCFs  $C_T(k,t)$, $M_2^T(k,t)$, $M_3^T(k,t)$ and $M_4^T(k,t)$ of these variables~\cite{Mokshin2015,Lee2000,Zwanzig2001}:
\begin{subequations}\label{15xy}
\begin{equation}
\frac{\partial^2}{\partial t^2} C_T(k,t) + \Delta_{T,\;2}^2(k)C_T(k,t) + \Delta_{T,\;3}^2(k)\int_{0}^{t} M_3^T(k,t-\tau ) \frac{\partial}{\partial \tau '} C_T(k,{\tau }') d\tau = 0, \label{15x}
\end{equation}
\begin{equation}
\frac{\partial^2}{\partial t^2} M_2^T(k,t) + \Delta_{T,\;3}^3(k)M_2^T(k,t) + \Delta_{T,\;4}^2(k)\int_{0}^{t} M_4^T(k,t-\tau ) \frac{\partial}{\partial \tau '} M_2^T(k,{\tau }') d\tau = 0. \label{15y}
\end{equation}
\end{subequations}
Applying the Laplace transformation
\begin{equation} \label{eq: time_scale_invariance}
LT\{f\}(s) = \tilde{f}(s) = \int_0^{\infty} \exp(-st) f(t) dt, \ \ \ s=i\omega,
\end{equation}
to Eqs. (\ref{15xy}) and expressing the Laplace image of the transverse current TCF $\widetilde{C}_T(k,s)$, we find
\begin{eqnarray}\label{eq_CT_rec2}
\widetilde{C}_T(k,s) =\cfrac{1}{s + \cfrac{\Delta_{T,\;2}^2(k)}{s + \cfrac{\Delta_{T,\;3}^2(k)}{s + \Delta_{T,\;4}^2(k) \widetilde{M}_4^T(k,s)}}} =\cfrac{1}{s + \cfrac{\Delta_{T,\;2}^2(k)}{s + \cfrac{\Delta_{T,\;3}^2(k)}{s + \cfrac{\Delta_{T,\;4}^2(k)}{s + \ddots }}}}.
\end{eqnarray}
The quantity $\widetilde{C}_T(k,s)$ is related to the spectral density $C_T(k,\omega)$ by
\begin{equation} \label{030}
C_T(k,\omega)=\frac{1}{\pi}\; \mathrm{Re} \left[\widetilde{C}_T(k,s = i\omega)\right].
\end{equation}

The key idea of a self-consistent relaxation theory of equilibrium liquids is that the description of the equilibrium dynamics in the whole range of wave numbers $k$ is realized using a dynamical variable set $\textbf{A}(k)$ related to the variables in the hydrodynamic theory~\cite{Mokshin_JPCM2018}. Namely, the dynamical variable set $\textbf{A}(k)$ either corresponds exactly to the hydrodynamic quantity or contains the hydrodynamic quantity in its expression (see, e.g., the discussion in Sec.~\ref{sec: Dyn_var})). In the case of describing transverse collective dynamics, the variables in the corresponding hydrodynamic equations are related to the dynamical variables $A_1^T(k)$ and $A_2^T(k)$.

In the hydrodynamic limit $k \to 0$, variables such as current and energy vary slowly with large relaxation times~\cite{Gotze2009}. This allows excluding other variables from consideration using a simple Markov approximation in the kinetic equations~\cite{Resibua/Lener}. In accordance with the self-consistent relaxation theory, typical relaxation times of processes related to the hydrodynamic variables and other variables also differ, and this difference can be significant. But the description is reduced not by neglecting ``nonhydrodynamic'' variables but by assuming the their characteristic time scales are comparable:
\begin{eqnarray}\label{34}
1/\sqrt{\Delta_{T,\;l}^2(k)} = 1/\sqrt{\Delta_{T,\;l+1}^2(k)} = 1/\sqrt{\Delta_{T,\;l+2}^2(k)}, \ \ \ l \in \mathbb{N}^*.
\end{eqnarray}
In case of transverse collective particle dynamics with the extended variable set containing the energy current $A_3^T(k)$ in addition to $A_1^T(k)$ and $A_2^T(k)$, we have $l=4$. We note that a completely \textit{equivalent} variable set (the density $A_0(k)$, longitudinal current $A_1(k)$, energy $A_2(k)$, and energy current $A_3(k)$ is realized in describing longitudinal collective particle dynamics in equilibrium liquids~\cite{Mokshin_JPCM2018}.

From~(\ref{eq_CT_rec2}) with (\ref{eq: time_scale_invariance}) taken into account, we obtain expressions for the Laplace images of the TCFs:
\begin{subequations} \label{eq: CT_k_omega}
\begin{eqnarray}\label{36}
\widetilde{M}_{3}^T(k,s)  = \frac{-s \pm \sqrt{ s^2 + 4\Delta_{T,\;4}^2(k) }}{2\Delta_{T,\;4}^2(k)},
\end{eqnarray}
\begin{equation}
\widetilde{M}_{2}^T(k,s) = \frac{2\Delta_{T,\;4}^2(k)}{s[2\Delta_{T,\;4}^2(k)-\Delta_{T,\;3}^2(k)] \pm \Delta_{T,\;3}^2(k)\sqrt{s^2 + 4\Delta_{T,\;4}^2(k)}},
\end{equation}
and
\begin{eqnarray} \label{eq: spectr_CT}
\widetilde{C}_{T}(k,s) &=& \widetilde{M}_{1}^T(k,s) =\\
&=& \frac{s[2\Delta_{T,\;4}^2(k)-\Delta_{T,\;3}^2(k)] \pm \Delta_{T,\;3}^2(k)\sqrt{s^2 + 4\Delta_{T,\;4}^2(k)}}{2\Delta_{T,\;2}^2(k)\Delta_{T,\;4}^2(k) + s^2 [2\Delta_{T,\;4}^2(k) - \Delta_{T,\;3}^2(k)] \pm s \Delta_{T,\;3}^2(k)\sqrt{s^2+4\Delta_{T,\;4}^2(k)}}. \nonumber
\end{eqnarray}
\end{subequations}
From~(\ref{eq: spectr_CT}) and (\ref{030}), we then directly obtain an expression for the spectral density of the transverse particle current:
\begin{subequations}
\begin{eqnarray} \label{eq: CT_omega}
C_T(k,\omega) = \frac{1}{2\pi}\; %\frac{k_B T}{ m}\;
\frac{\Delta_{T,\;2}^2(k)\Delta_{T,\;3}^2(k)}{\Delta_{T,\;4}^2(k)-\Delta_{T,\;3}^2(k)}\;\frac{\pm \sqrt{4\Delta_{T,\;4}^2(k)-\omega^2}}{\omega^4 + \mathcal{A}_1^T(k)\omega^2 + \mathcal{A}_2^T(k)},
\end{eqnarray}
where
\begin{eqnarray}\label{eq_CT_theory_a}
\mathcal{A}_1^T(k) = \frac{\Delta_{T,\;3}^4(k)+\Delta_{T,\;2}^2(k)\Delta_{T,\;3}^2(k)-2\Delta_{T,\;2}^2(k)\Delta_{T,\;4}^2(k)}{\Delta_{T,\;4}^2(k)-\Delta_{T,\;3}^2(k)},
\end{eqnarray}
\begin{eqnarray}\label{eq_CT_theory_b}
\mathcal{A}_2^T(k) = \frac{\Delta_{T,\;2}^4(k)\Delta_{T,\;4}^2(k)}{\Delta_{T,\;4}^2(k)-\Delta_{T,\;3}^2(k)}.
\end{eqnarray}\label{eq_CT_theory}
\end{subequations}

%~~~~~~~~~~~~~~~~~~~~~~~~figure~~~~~~~~~~~~~~~~~~~~~~~~~~~~~~~~~~~~~~~~~~~~
\begin{figure}
\begin{center}
\includegraphics[height=8cm,angle=0]{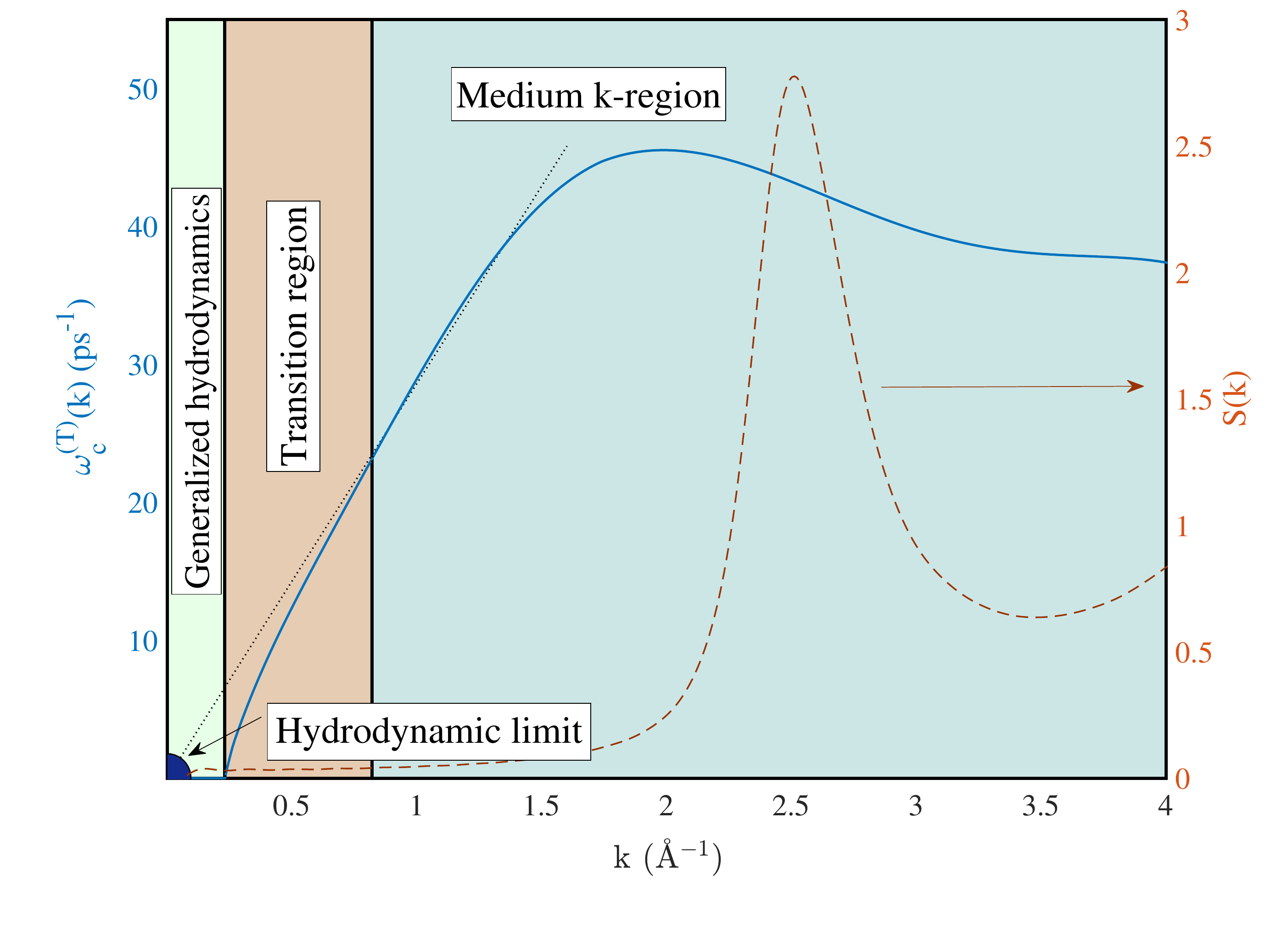} \vskip -0.8cm
\includegraphics[height=10cm,angle=0]{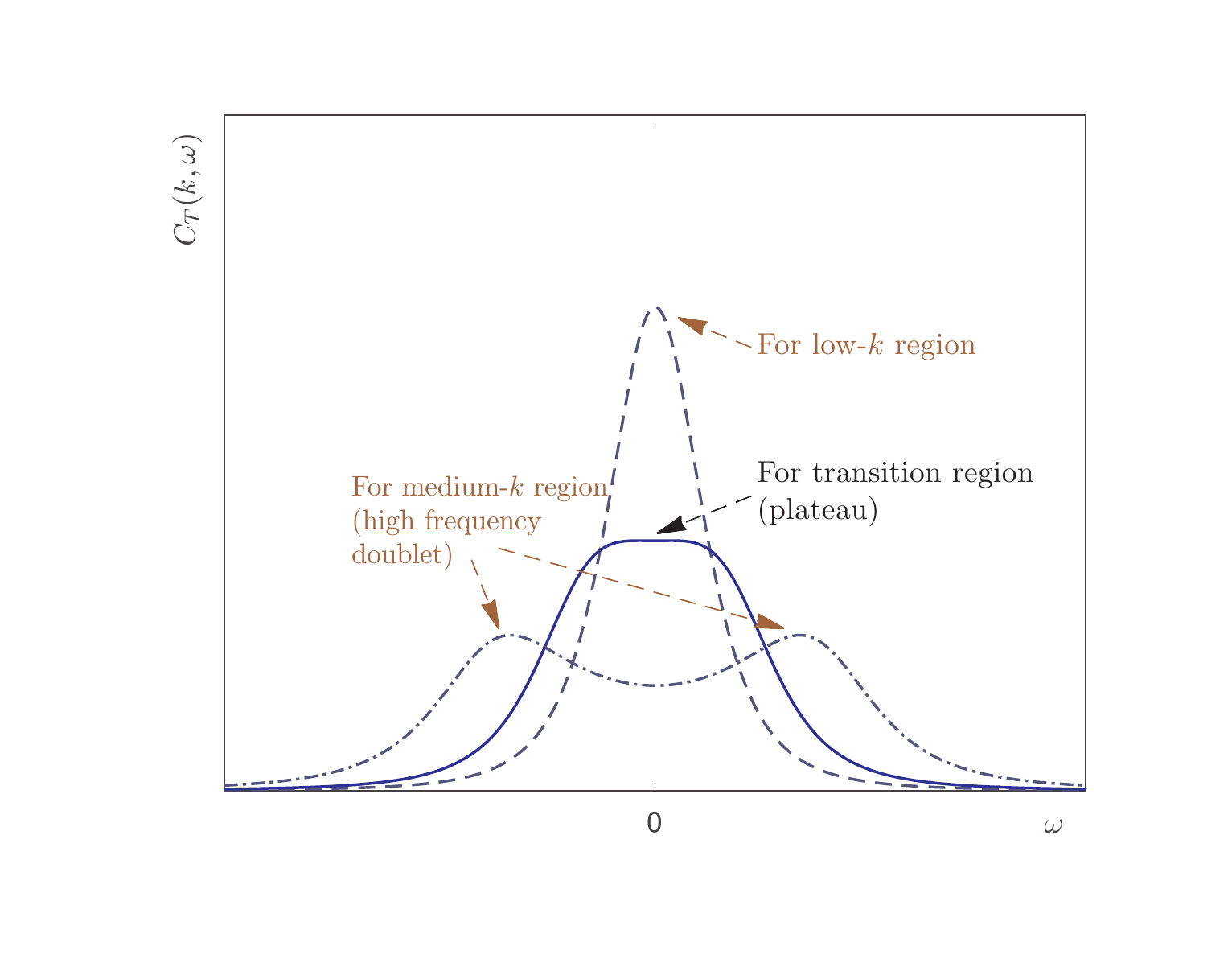} \vskip -0.8cm
\caption{\textbf{Top:} Schematic illustrating the dispersion law of similar transverse acoustic oscillations $\omega_c^{(T)}(k)$ and the static structure factor $S(k)$ of some equilibrium liquid. \textbf{Bottom:} Schematic showing the form of the spectrum of the transverse current $C_{T}(k,\omega)$ for the domain of intermediate values of the wave number $k$, for transition regime, and for the domain of generalized hydrodynamics (the small-$k$ limit).}
\label{Fig_01}
\end{center}
\end{figure}
%~~~~~~~~~~~~~~~~~~~~~~~~figure~~~~~~~~~~~~~~~~~~~~~~~~~~~~~~~~~~~~~~~~~~~~

As follows from~(\ref{eq_CT_theory}), the contour $C_T(k,\omega)$ of the frequency spectrum is determined by the frequency parameters $\Delta_{T,\;2}^2(k)$, $\Delta_{T,\;3}^2(k)$ and $\Delta_{T,\;4}^2(k)$ and is consequently determined by the structural features of the system and the interparticle interaction and also depends on the thermodynamic state of the system.

We can distinguish characteristic regimes in the transverse collective particle dynamics of an equilibrium liquid, which are conveniently considered in correspondence with the dispersion law of similar transverse acoustic oscillations $\omega_c^{(T)}(k)$ and with a static structure factor $S(k)$ (see Fig.~\ref{Fig_01} (top)). We have the following c regimes as the wave number increases on the $k$ scale: the hydrodynamic regime ($k \to 0$), generalized hydrodynamics, the transition regime, the domain of intermediate values of $k$, and the short-wave limit (not shown in Fig.~\ref{Fig_01}).

\vskip 0.5cm

\textbf{Short-wave limit.}  The particle dynamics in the short-wave limit is characterized by a single parameter, the mean particle velocity $\bar{v}_x$.  The frequency relaxation parameters are related by the recurrence relation
\begin{equation} \label{eq: recurr}
\Delta_{T,\;l+2}^2(k) = (l+1)\; \Delta_{T,\;2}^2(k), \ \ \ l=1,\;2,\; \ldots,
\end{equation}
where the equality
\[
\Delta_{T,\;2}^2(k) = (k \bar{v}_x)^2
\]
follows from (\ref{eq: Delta1_micro}) if we neglect the interparticle interaction. In this case, expression (\ref{eq_CT_theory}) for the spectral density $C_T(k,\omega)$ with~(\ref{eq: recurr}) taken into account becomes
\begin{equation}
C_T(k,\omega) =  \sqrt{ \frac{1}{2\pi \; \Delta_{T,\;2}^2(k)} }\; \exp{\left ( - \frac{\omega^2}{2 \Delta_{T,\;2}^2(k)} \right )},
\end{equation}
which corresponds exactly to the correct model result~(\ref{eq: spectral_density_CT}).

\vskip 0.5cm

\textbf{Domain of intermediate values of \textit{k}.}  The domain of intermediate values of $k$ is realized for wave numbers corresponding to spatial scales comparable to the effective action scale of the interaction potential between particles. This wave-number domain is shown in dark gray in Fig.~\ref{Fig_01} (top). At these spatial scales, shear stiffness appears in a liquid, and quasi-solid-state collective dynamics supporting transverse oscillatory processes are realized. In this regime, the spectral density $C_T(k,\omega)$ is a high-frequency doublet: two expressed symmetric maximums located at the frequencies $\pm \; \omega_c^{(T)}$ and a minimum at the frequency zero ($\omega = 0$) (Fig.~\ref{Fig_01} (bottom)). Moreover, the dispersion $\omega_c^{(T)} = \omega_c^{(T)}(k)$ of transverse collective excitations in an equilibrium high-density liquid near the melting point must be extrapolated to the domain of small wave numbers in the form of the usual linear relation
\begin{equation}
\omega_c^{(T)}(k) = v^{(T)} k,
\end{equation}
where $v^{(T)}$ is propagation velocity of transverse sound waves.

The frequency relaxation parameter $\Delta_{T,\;4}^2(k)$ characterizes the minimum time scale $\tau_{T,\;4}(k) = 1/\sqrt{\Delta_{T,\;4}^2(k)}$ at which the transverse collective particle dynamics is realized. Consequently, for the frequency domain in which we are interested, we can write
\begin{equation}
\label{eq: cond1}
\omega^2 \ll \Delta_{T,\;4}^2(k).
\end{equation}
Expression (\ref{eq_CT_theory}) then becomes
\begin{equation} \label{eq: CT_mediate_k}
C_T(k,\omega) = \frac{1}{\pi}\; %\frac{k_B T}{ m}\;
\frac{\Delta_{T,\;2}^2(k)\Delta_{T,\;3}^2(k) \sqrt{\Delta_{T,\;4}^2(k)}}{\Delta_{T,\;4}^2(k)-\Delta_{T,\;3}^2(k)}\;\frac{1}{\omega^4 + \mathcal{A}_1^T(k)\omega^2 + \mathcal{A}_2^T(k)},
\end{equation}
where the coefficients $\mathcal{A}_1^T(k)$ and $\mathcal{A}_2^T(k)$ are defined by (\ref{eq_CT_theory_a}) and (\ref{eq_CT_theory_b}).

We note several important properties of expression~(\ref{eq: CT_mediate_k}).
First, because the spectral density can take
only nonnegative values, $C_T(k,\omega) \geq 0$, the relation between the frequency parameters $\Delta_{T,\;3}^2(k)$ and $\Delta_{T,\;4}^2(k)$
\begin{equation}
\label{eq: cond0}
\Delta_{T,\;4}^2(k) \geq \Delta_{T,\;3}^2(k),
\end{equation}
must be satisfied, which completely agrees with condition~(\ref{eq: cond1}) for the minimum time scale.

Second, for a fixed $k$, the form of the spectrum $C_T(k,\omega)$ is defined by the biquadratic polynomial in the denominator in~(\ref{eq: CT_mediate_k}). Almost all features of the spectrum $C_T(k,\omega)$ can be rather easily determined in the framework of the usual mathematical analysis of~(\ref{eq: CT_mediate_k}).

Further, solving the problem of the existence of extremums, we obtain the dispersion law from~(\ref{eq: CT_mediate_k}):
\begin{subequations}
\begin{equation}
\label{eq_omegaT_1}
\omega_c^{(T)}(k)=\sqrt{-\frac{\mathcal{A}_{1}^T(k)}{2}}
\end{equation}
or
\begin{equation}
\label{eq_omegaT_2}
\omega_c^{(T)}(k) = \sqrt{\Delta_{T,\;2}^{2}(k)  - \frac{ \Delta_{T,\;3}^{2}(k)[\Delta_{T,\;3}^2(k)-\Delta_{T,\;2}^2(k)]}{2[\Delta_{T,\;4}^{2}(k)-\Delta_{T,\;3}^{2}(k)]}}.
\end{equation}
\end{subequations}
As can be seen from (\ref{eq_omegaT_1}), the high-frequency doublet in the spectrum $C_T(k,\omega)$ appears under the condition $\mathcal{A}_{1}^T(k)<0$ [see Fig.~\ref{Fig_01}], which is similar to the existence condition for high-frequency excitations in accordance with the viscoelastisity model (see p.~268 in~\cite{Hansen/McDonald}):
\begin{equation} \label{eq: disp_viscoelastic}
\omega_c^{(T)}(k) = \sqrt{\Delta_{T,\;2}^2(k) - \frac{1}{2\tau_{t}^2(k)}}
\end{equation}
with the relaxation time  $\tau_t(k)$. Comparing (\ref{eq_omegaT_2}) and (\ref{eq: disp_viscoelastic}), we find the definition of the relaxation time in the framework of the self-consistent relaxation theory with expression (\ref{eq: CT_mediate_k}):
\begin{equation}
\tau_t^2(k) = \frac{\Delta_{T,\;3}^{2}(k)[\Delta_{T,\;3}^2(k)-\Delta_{T,\;2}^2(k)]}{\Delta_{T,\;4}^{2}(k)-\Delta_{T,\;3}^{2}(k)}.
\end{equation}
It is important that the quantity $\tau_t^2(k)$ remains undefined in the viscoelasticity model. An interesting consequence of the obtained expression~(\ref{eq_omegaT_1}) for the dispersion law that the dispersion of transverse collective excitations is determined by the interaction potential $u(r)$, between particles and also by two-, three-, and four-particle correlations. This conclusion follows directly from the microscopic expressions for the frequency relaxation parameters in (\ref{eq_omegaT_1}).

Finally, expression~(\ref{eq: CT_mediate_k}) allows estimating the values of physical quantities characterizing the propagation of shear oscillations in a liquid: the propagation velocity $v^{(T)}$ and the damping coefficient $\Gamma^{(T)}$ of transverse sound oscillations. Hence, for the velocity $v^{(T)}$, from dispersion relation~(\ref{eq_omegaT_2}), we obtain
\begin{subequations}
\begin{equation}
\label{eq: v_T_1}
v^{(T)} k = \lim_{k \to 0}  \sqrt{\Delta_{T,\;2}^{2}(k)  - \frac{ \Delta_{T,\;3}^{2}(k)[\Delta_{T,\;3}^2(k)-\Delta_{T,\;2}^2(k)]}{2[\Delta_{T,\;4}^{2}(k)-\Delta_{T,\;3}^{2}(k)]}}
\end{equation}
or
\begin{equation}
\label{eq: v_T_2}
v^{(T)} k = \lim_{k \to 0} \sqrt{\frac{2\Delta_{T,\;2}^2(k) \Delta_{T,\;4}^2(k)}{2\Delta_{T,\;4}^2(k)-\Delta_{T,\;3}^2(k)}}.
\end{equation}
\end{subequations}
Expression~(\ref{eq: v_T_2}) is obtained as a result of solving the dispersion equation solution according to the approximation scheme proposed by Mountain in~\cite{Mountain_1966} (also see \cite{Mokshin_JPCM2018}). Expressions~(\ref{eq: v_T_1}) and (\ref{eq: v_T_2}) must give close values for $v^{(T)}$. For the damping coefficient of transverse sound, we obtain
\begin{eqnarray}
\label{eq: Gamma_T}
\Gamma^{(T)} k^2 = \lim_{k \to 0} \; \frac{\Delta_{T,\;3}^2(k)\sqrt{\Delta_{T,\;4}^2(k)}}{2\Delta_{T,\;4}^2(k)-\Delta_{T,\;3}^2(k)}.
\end{eqnarray}

Therefore, we can respectively compute the values of $v^{(T)}$ and $\Gamma^{(T)}$ from~(\ref{eq: v_T_1}) or (\ref{eq: v_T_2}) and~(\ref{eq: Gamma_T}). For these computations, we must use values of the frequency parameters corresponding to wave numbers in the domain of intermediate values of $k$, where we have quasi-solid-state properties of the liquid. We note that~(\ref{eq: Gamma_T}) is also interesting because direct experimental measurements of $\Gamma^{(T)}$ are rather complicated.

\vskip 0.5 cm

\textbf{Transition regime}. Expression (\ref{eq: CT_mediate_k}) for the spectral density $C_{T}(k,\omega)$ with a dispersion law $\omega_c^{(T)}(k)$ of form ~(\ref{eq_omegaT_1}) indicates the existence of regime in the transverse collective particle dynamics, where we have a transition from quasi-solid-state dynamics, discussed above, to the dynamics of an ordinary equilibrium liquid. In Fig~\ref{Fig_01} (top), the wave number domain corresponding to this regime is shown in gray. Here, the high-frequency doublet in the spectrum $C_{T}(k,\omega)$ is transformed into a single broadened spectral component with the form of smoothed trapezoid centered on the frequency $\omega = 0$ (see Fig.~\ref{Fig_01} (bottom)). The tangents to the dispersion curve $\omega_c^{(T)}(k)$ for the corresponding wave numbers are extrapolated to finite values $k$ as $\omega_c^{(T)} \to 0$ (see Fig.~\ref{Fig_01} (top)).

The existence of this regime is quite understandable because the transition to quasi-solid-state dynamics occurs in some domain of spatial scales and consequently in some wave number domain. An analysis of expression (\ref{eq: CT_mediate_k}) for $C_{T}(k,\omega)$ shows that such a transition is realized for the correspondence between frequency relaxation parameters
\begin{eqnarray}\label{eq_main_conditions}
\Delta_{T,\;3}^{2}(k)&=&3\Delta_{T,\;2}^{2}(k), \\
\Delta_{T,\;4}^{2}(k)&=&2\Delta_{T,\;3}^{2}(k) = 6\Delta_{T,\;2}^{2}(k). \nonumber
\end{eqnarray}
In this case, for the coefficients $\mathcal{A}_1^T(k)$ and $\mathcal{A}_2^T(k)$ in~(\ref{eq: CT_mediate_k}), we obtain
\begin{subequations}
\begin{equation} \label{eq_CT_theory_trans_a}
\mathcal{A}_1^T(k) = 0,
\end{equation}
\begin{equation} \label{eq_CT_theory_trans_aa}
\mathcal{A}_2^T(k) = 2\Delta_{T,\;2}^4(k),
\end{equation}
\end{subequations}
and expression~(\ref{eq: CT_mediate_k}) becomes
\begin{eqnarray}\label{eq_CT_theory_trans}
C_T(k,\omega) = \frac{1}{\pi} \; %\frac{k_B T}{ m}\;
\frac{\Delta_{T,\;2}^2(k)\sqrt{6\Delta_{T,\;2}^2(k)}}{\omega^4 + 2\Delta_{T,\;2}^4(k)}.
\end{eqnarray}
In fact, correspondence law~(\ref{eq_main_conditions}) for the frequency relaxation parameters is equivalent to relation~(\ref{eq_CT_theory_trans_a}) and represents existence condition for this transition mode.

In turn, as can be seen from~(\ref{eq_CT_theory_trans}), transverse collective dynamics is determined by only the frequency relaxation parameter ~$\Delta_{T,\;2}^2(k)$ and consequently by the interaction potential $u(r)$ between particles, the particle radial distribution function $g(r)$, and the mean particle velocity $\bar{v}_x$. This regime is related to the transition from the collective dynamics description in terms of microscopic characteristics to the description where hydrodynamics parameters such as the generalized viscosity $\nu(k)$ and the thermal conductivity $D_T(k)$ can be used.

\vskip 0.5 cm
	
\textbf{Generalized hydrodynamics regime}. The generalized hydrodynamics regime corresponds to small wave numbers and passes into the usual hydrodynamic regime in the limit $k \to 0$. Here, we have typical liquid collective dynamics, and transverse acoustic waves are absent, i.e., $\omega_c^{(T)}=0$. In Fig.~\ref{Fig_01} (top), the wave number domain corresponding to this regime is shown in light gray. Here, as in the case of the hydrodynamic regime, the spectrum $C_T(k,\omega)$ has only a single component at the frequency zero and is reproduced by a Lorentz function (see expression~(\ref{eq: hydro3}) and Fig.~\ref{Fig_01} (bottom)). The width of this spectral component is determined by the generalized shear viscosity $\nu(k)$, whose values depend on the wave number $k$. In passing to the hydrodynamic limit $k \to 0$, $\nu(k)$ is extrapolated to the hydrodynamic kinematic shear viscosity,
\[
\nu = \lim_{k \to 0} \; \nu(k).
\]

In this regime, the transverse collective particle dynamics is described by the hydrodynamic variables $A_1^T(k)$ and $A_2^T(k)$. Consequently, the equalization of the time scales given by relation~(\ref{34}) occurs for $l=3$:
\begin{eqnarray}\label{eq: delta3_4}
1/\sqrt{\Delta_{T,\;3}^2(k)} = 1/\sqrt{\Delta_{T,\;4}^2(k)} = 1/\sqrt{\Delta_{T,\;5}^2(k)} = \ldots,
\end{eqnarray}
and this time scale equality naturally realizes the transition from the extended variable set $A_1^T(k)$, $A_2^T(k)$ and $A_3^T(k)$ to the usual hydrodynamic variables $A_1^T(k)$ and $A_2^T(k)$. Taking (\ref{eq: delta3_4}) into account, from (\ref{eq: CT_omega}), we obtain
\begin{subequations}
\begin{eqnarray}\label{eq_CT_theory_gp}
C_T(k,\omega) = \frac{1}{\pi}\; %\frac{k_B T}{ m}\;
\frac{\Delta_{T,\;2}^2(k)\sqrt{\Delta_{T,\;3}^2(k)}}{\Delta_{T,\;3}^2(k)-\Delta_{T,\;2}^2(k)} \; \frac{1}{\omega^2 + \mathcal{A}_2^T(k)},
\end{eqnarray}
where
\begin{equation}
\mathcal{A}_2^T(k)=\frac{\Delta_{T,\;2}^4(k)}{\Delta_{T,\;3}^2(k)-\Delta_{T,\;2}^2(k)}.
\end{equation}
\end{subequations}

Expression~(\ref{eq_CT_theory_gp}) is characterized by the following properties. First, this expression determines the spectral density~$C_T(k,\omega)$ in the frequency domain
\begin{equation}
\label{eq: cond2}
\omega^2 \ll \Delta_{T,\;3}^2(k),
\end{equation}
where $\tau_{T,\;3}(k)=1/\sqrt{\Delta_{T,\;3}^2(k)}$ is the minimum time scale in the collective particle dynamics for wave numbers corresponding to the generalized hydrodynamics.

Second, expression~(\ref{eq_CT_theory_gp}) yields the correspondence between the frequency parameters
\begin{equation}
\Delta_{T,\;3}^2(k) \geq \Delta_{T,\;2}^2(k),
\end{equation}
which is analogous to relation~(\ref{eq: cond0}) and which just like condition~(\ref{eq: cond2}) reflects the fact that the existence of oscillatory processes related to transverse collective particle dynamics in the generalized hydrodynamic regime are characterized by frequencies not exceeding $\sqrt{\Delta_{T,\;3}^2(k)}$.

Third, expression~(\ref{eq_CT_theory_gp}) corresponds to the known hydrodynamic expression~(\ref{eq: hydro3}) for the spectral density $C_T(k,\omega)$ of the of transverse current, and coefficient $\mathcal{A}_2^T(k)$ is related to the generalized shear viscosity $\nu(k)$:
\begin{equation} \label{eq: shear_visc}
\nu(k) = \frac{1}{k^2}\sqrt{\mathcal{A}_2^T(k)} = \frac{1}{k^2} \frac{\Delta_{T,\;2}^2(k)}{\sqrt{\Delta_{T,\;3}^2(k)-\Delta_{T,\;2}^2(k)}}.
\end{equation}

Fourth, expression~(\ref{eq_CT_theory_gp}) becomes identical to hydrodynamic expression~(\ref{eq: hydro3}) under the relation between
the frequency parameters
\begin{equation}
\Delta_{T,\;3}^2(k) \gg \Delta_{T,\;2}^2(k),
\end{equation}
which is equivalent to the assumption that the time scales are separated and to the Markov approximation, which are used in hydrodynamic theory~\cite{Hansen/McDonald} and lead to Eq.~(\ref{eq: hydro2}) and expression~(\ref{eq: hydro3}) for $C_T(k,\omega)$.

\section{Transverse collective dynamics of liquid lithium near the melting point \label{sec: comparison}}
	
A self-consistent relaxation theory of longitudinal collective particle dynamics in simple liquids was presented in~\cite{Mokshin_JPCM2018}. As an example, the theoretically calculated spectra of the dynamic structure factor $S(k,\omega)$ and the dispersion law $\omega_c^{(L)}(k)$ of longitudinal sound waves were compared with experimental data on inelastic x-ray scattering in liquid lithium at the temperature $T=475$\;K (the melting temperature is $T_m=453.65$~K)~\cite{Scopigno2000}. Moreover, in the framework of the theory, the sound propagation velocity $v^{(L)}=6000 \pm 1200$~m/s, sound damping coefficient $\Gamma^{(L)}= 18.4 \pm 3.5$~nm$^2$/ps and thermal conductivity $D_T = 21.5 \pm 3.5$~nm$^2$/ps were computed. Here, we also quantitatively compute various characteristics of the transverse collective dynamics for liquid lithium at $T=475$\;K and the number density of atoms $\rho=0.0\,445$~$\textrm{\AA}^{-3}$. We compare the theoretical results with the results of the equilibrium atomic dynamics simulations in an $NVT$ ($N$ is the number of particles, $V$ is the volume, $T$ is the temperature)-ensemble~\cite{Murtazayev1999}.

The modeling details are analogous to those presented in~\cite{Mokshin_JPCM2018,Khusnutdinoff2018}. We performed the simulation calculations for a system of $N=16\,000$ atoms enclosed in cubic box with periodic boundary conditions. The interaction between lithium atoms was given by a spherical pseudopotential~\cite{Gonzalez1993}. As we previously showed \cite{Khusnutdinoff2018}, this model of the potential agrees best with experimental data on elastic and inelastic x-ray scattering compared with known multiparticle potentials of the EAM (embedded atom model) type. The equations of motion were integrated according to the velocity Verlet algorithm with the time step $\tau '=0.01$~ps \cite{Mokshin2012}. To bring the system to the thermodynamic equilibrium and to calculate the spectral characteristics, we respectively performed $200\;000$ and $100\;000$ time steps.
	
%~~~~~~~~~~~~~~~~~~~~~~~~figure~~~~~~~~~~~~~~~~~~~~~~~~~~~~~~~~~~~~~~~~~~~~
\begin{figure*}
\begin{center}
%\hskip -5 cm
\includegraphics[height=10.0cm, angle=0]{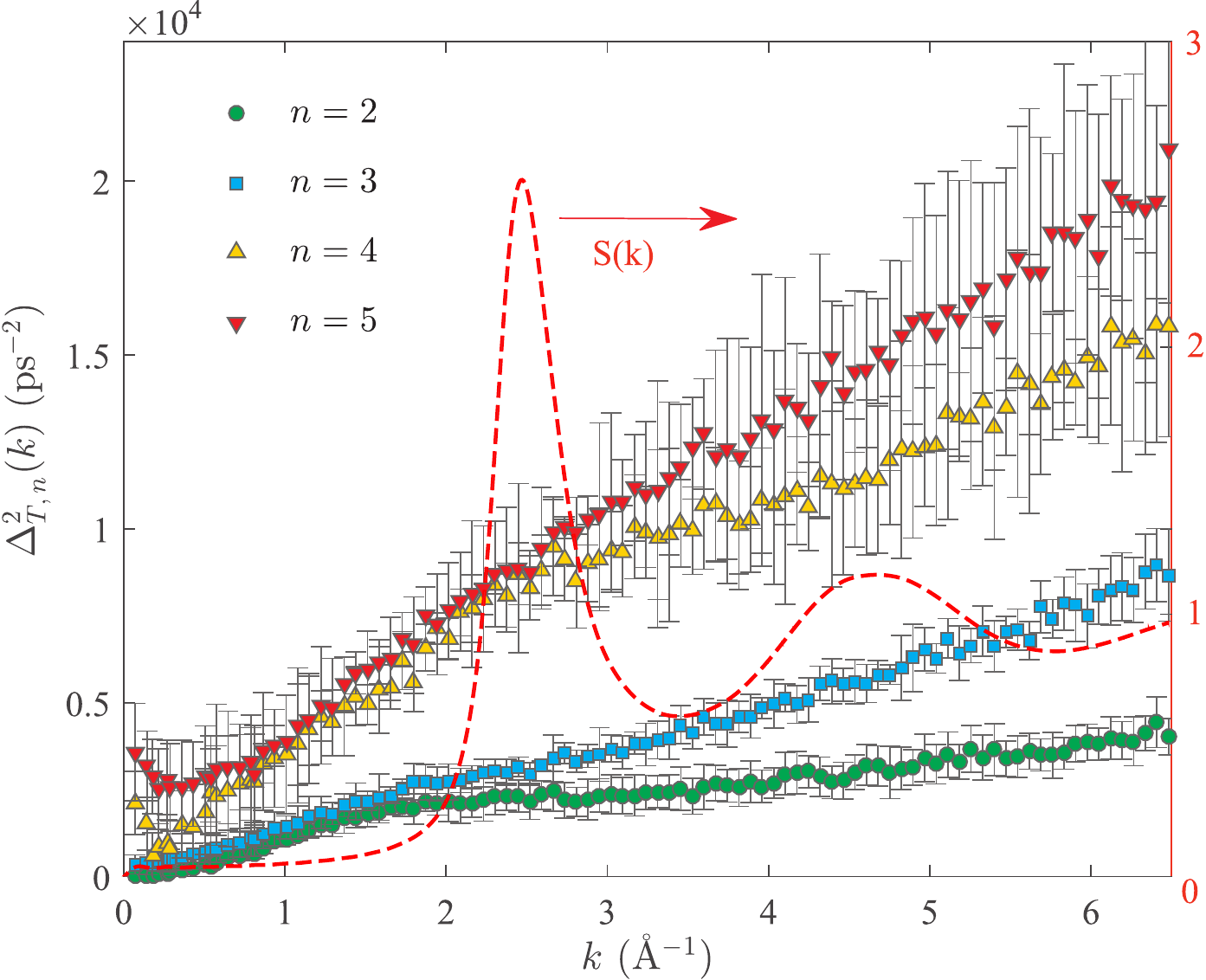}
\caption{Frequency relaxation parameter $\Delta_{T,\;2}^2(k)$, $\Delta_{T,\;3}^2(k)$, $\Delta_{T,\;4}^2(k)$ and $\Delta_{T,\;5}^2(k)$ of the transverse collective atom dynamics in equilibrium melted lithium at the temperature $T=475$\;K: the results were obtained based on configurational data for the atomic dynamics simulated. Vertical closed intervals denote the confidence intervals obtained as a result of numerical calculations with the configurational data at different instants. The dashed line is the experimental static structure factor $S(k)$ lithium at the given temperature~\cite{Waseda}.}
\label{Fig_02}
\end{center}
\end{figure*}
%~~~~~~~~~~~~~~~~~~~~~~~~figure~~~~~~~~~~~~~~~~~~~~~~~~~~~~~~~~~~~~~~~~~~~

%~~~~~~~~~~~~~~~~~~~~~~~~figure~~~~~~~~~~~~~~~~~~~~~~~~~~~~~~~~~~~~~~~~~~~~
\begin{figure*}
\begin{center}
\includegraphics[height=11cm, angle=0]{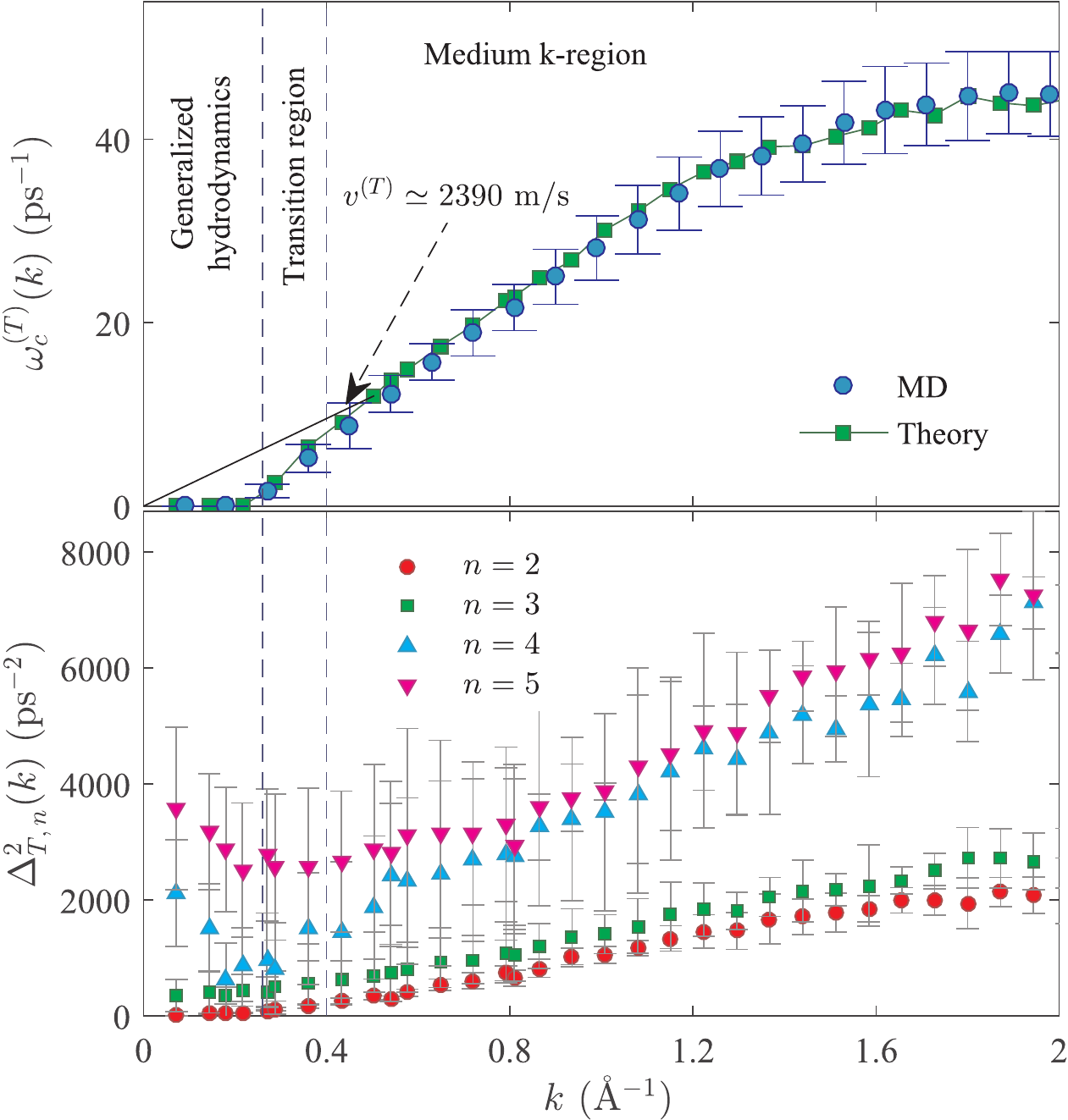}
\caption{\textbf{Top}: Dispersion law of transverse oscillatory dynamics of atoms of equilibrium melted lithium
at the temperature $T=475$~K: circles ($\circ ~\circ ~\circ$) denote data from atomic dynamics simulations data; squares joined by a solid line are the results of theoretical calculations with Eq. (\ref{eq_omegaT_2}); the straight line (T) (T) in the small wave number range is the extrapolated hydrodynamic result $\omega_c^{(T)}(k)=v^{(T)}k$, where $v^{(T)}=2\,390$~m/s is the transverse speed of sound.
\textbf{Bottom}: An enlarged fragment of Fig.~\ref{Fig_02} showing the frequency relaxation parameters $\Delta_{T,\;2}^2(k)$, $\Delta_{T,\;3}^2(k)$, $\Delta_{T,\;4}^2(k)$ and $\Delta_{T,\;5}^2(k)$ in the range of small finite
wave numbers: vertical dashed lines separate different regimes of the oscillatory dynamics.}
\label{Fig_03}
\end{center}
\end{figure*}
%~~~~~~~~~~~~~~~~~~~~~~~~figure~~~~~~~~~~~~~~~~~~~~~~~~~~~~~~~~~~~~~~~~~~~

Based on the configurational data representing an array data structure of atom coordinates and velocities at an arbitrary fixed instant, we calculated frequency moments and also frequency relaxation parameters $\Delta_{T,\;n}^{2}(k)$, $n=2$, $3$, $4$ and $5$, in a wide range of wave numbers $k \in [0.072,\; 6.481]$~\AA$^{-1}$ [see expressions~(\ref{eq: freq_parameters}) and (\ref{eq: Delta_vs_moments})]. We note that basic definition~(\ref{eq: freq_parameters}) and microscopic expression~(\ref{eq: Delta1_micro}) for the frequency parameter $\Delta_{T,\;2}^2(k)$ give identical results. As can be seen from Fig.~\ref{Fig_02}, the calculated frequency parameters have a similar monotonic dependence on $k$ in the wave number range $k>0.25$~\AA$^{-1}$. Just as expected, the typical time scales $\tau_{T,\;n}(k) = 1/\sqrt{\Delta_{T,\;n}^2(k)}$ shorten as $k$ increases. The minimum values are taken by $\tau_{T,\;4}(k) = 1/\sqrt{\Delta_{T,\;4}^2(k)}$ and $\tau_{T,\;5}(k) = 1/\sqrt{\Delta_{T,\;5}^2(k)}$, determining time scales less than $0.01$\;ps. Such time scales are even shorter than the step $\tau'$ for integrating the atomic equation of motion in the simulation calculations. The time $\tau_{T,\;4}(k)$ can therefore indeed be regarded as a threshold quantity in the hierarchy of time scales for the relaxation process~\cite{Bogolyubov_book}. In the wave number range $k \leq 3.5$~\AA$^{-1}$, the frequency parameters $\Delta_{T,\;4}^2(k)$ and $\Delta_{T,\;5}^2(k)$ take comparable values and show a difference only as $k$ increases further, approaching correspondence (\ref{eq: recurr}) characteristic of the free-moving particle regime.

%~~~~~~~~~~~~~~~~~~~~~~~~figure~~~~~~~~~~~~~~~~~~~~~~~~~~~~~~~~~~~~~~~~~~~~
\begin{figure*}
\begin{center}
\includegraphics[width=14cm, angle=0]{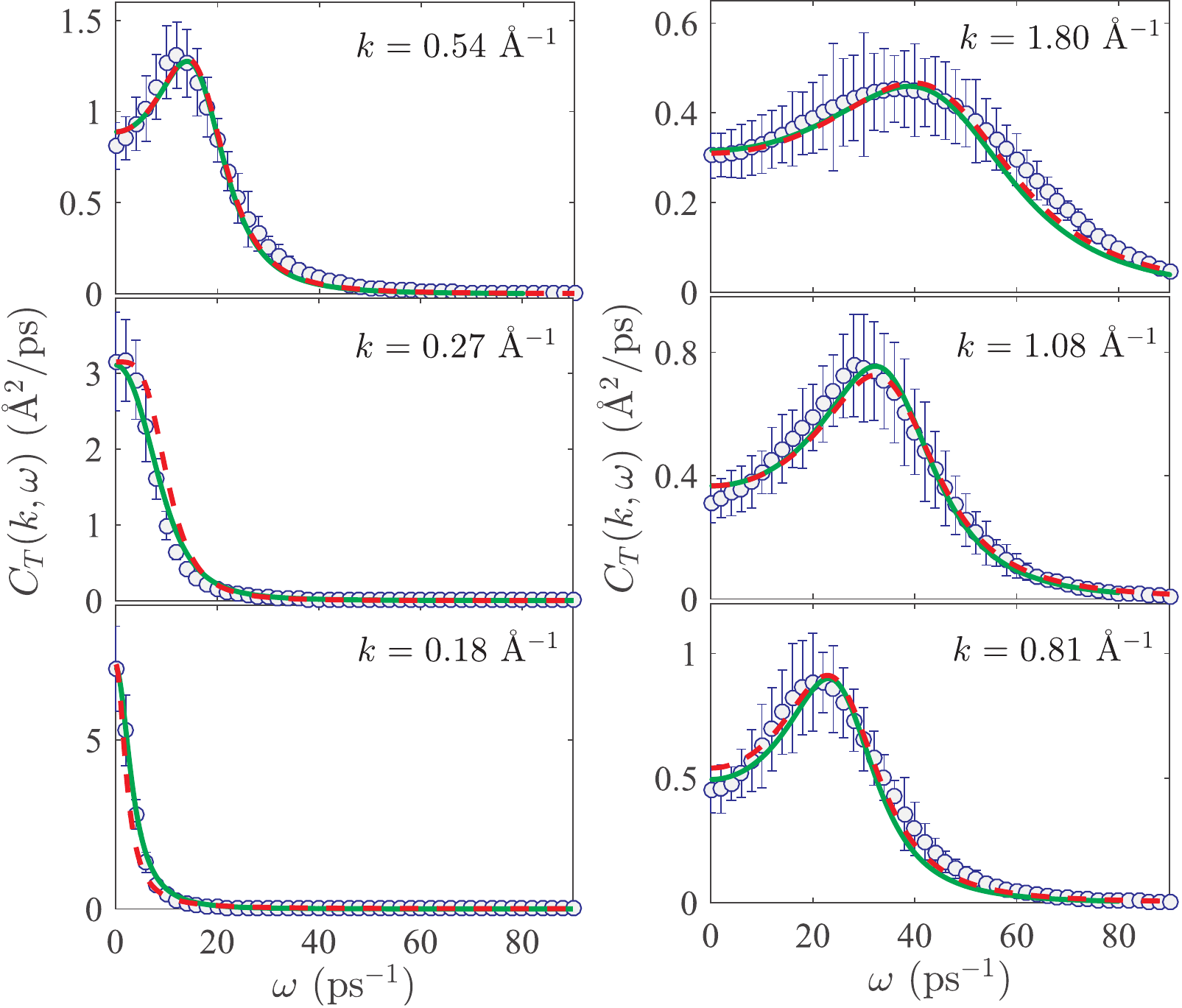}
\caption{Spectra ${C}_T(k,\omega)$ of transverse current of melted lithium at the temperature $T=475$~K for
different wave numbers: circles ($\circ ~\circ ~\circ$)  are denote the atomic dynamics simulations data; the solid line is the
theoretical results with general expression~(\ref{eq_CT_theory}); the dotted line shows theoretical results for different
regimes, Eq. (\ref{eq_CT_theory_gp}) for $k=0.18$~\AA$^{-1}$;  Eq. (\ref{eq_CT_theory_trans}) for $k=0.27$~\AA$^{-1}$, and Eq. (\ref{eq: CT_mediate_k}) for $k=0.54$~\AA$^{-1}$, $0.81$~\AA$^{-1}$, $1.08$~\AA$^{-1}$ and $1.80$~\AA$^{-1}$.}
\label{Fig_04}
\end{center}
\end{figure*}
%~~~~~~~~~~~~~~~~~~~~~~~~figure~~~~~~~~~~~~~~~~~~~~~~~~~~~~~~~~~~~~~~~~~~~

The obtained frequency relaxation parameter values allow calculating the dispersion relation $\omega_c^{(T)}(k)$ of the transverse oscillatory dynamics based on Eq. (\ref{eq_omegaT_2}). As can be seen from Fig.~\ref{Fig_03}, the theoretical expression for the dispersion relation completely reproduces the modeling results. Moreover, the typical regimes of collective oscillatory dynamics in melted lithium at the given temperature can be clearly seen in the plot of $\omega_c^{(T)}(k)$. The \textit{generalized hydrodynamics regime} corresponds to small $0.25$~\AA$^{-1} < k < 0.4$~\AA$^{-1}$, the \textit{transition regime} is realized on the interval $k \geq 0.4$~\AA$^{-1}$, and the \textit{domain of intermediate values} of $k$, where shear oscillatoty processes occur, is $k \geq 0.4$~\AA$^{-1}$.
	
In Fig.~\ref{Fig_04}, we compare the spectra $C_T(k,\omega)$ obtained based on the atomic dynamics simulations data and calculated according to general expression~(\ref{eq_CT_theory}). Here, we also give the theoretical spectra $C_T(k,\omega)$ calculated by formulas (\ref{eq_CT_theory_gp}), (\ref{eq_CT_theory_trans}) and (\ref{eq: CT_mediate_k}) or the corresponding typical collective atomic dynamics regimes of lithium. As can be seen from the figure, the theoretical and modeling results agree completely. The theory reproduces all features of the spectra $C_T(k,\omega)$ for different wave numbers, correctly reproducing the transition from dynamics the expressed high-frequency oscillation modes [see the spectra $C_T(k,\omega)$ in Fig.~\ref{Fig_04}] for wave numbers from $k=0.54$~\AA$^{-1}$ to $k=1.80$~\AA$^{-1}$] to the usual transverse collective dynamics for a liquid [see the spectrum $C_T(k,\omega)$ at $k=0.18$~\AA$^{-1}$ in Fig.~\ref{Fig_04}].

Based on Eqs.~(\ref{eq: v_T_2}) and (\ref{eq: Gamma_T}), we determined transverse speed of sound $v^{(T)} \simeq 2\,390$~m/s and the damping coefficient $\Gamma^{(T)} \simeq 17$~\AA$^2$/ps of the transverse sound mode. Numerical calculations with Eq.~(\ref{eq: shear_visc}) show that the generalized kinematic shear viscosity depends on $k$ as
\begin{equation}
\nu(k) = \frac{\nu}{1+ \alpha^2 k^2},
\end{equation}
where the coefficient $\alpha$ and kinematic shear viscosity $\nu$ take the values $\alpha=1.1$ and  $\nu = 105 \pm 11$~\AA$^2$/ps. It is remarkable that the obtained viscosity value agrees well with the experimental value: $\nu_{exp} = 111$~\AA$^2$/ps~\cite{link_1}.

\section{Concluding commentary}
In conclusion, we note the following points.

The presented self-consistent relaxation theory of transverse collective particle dynamics in equilibrium liquids is a realization of Bogoliubov's idea about a reduced description of relaxation processes in liquids. This theory is completely consistent with the theoretical description of longitudinal collective dynamics in liquids, which was previously given in detail in~\cite{Mokshin2015,Mokshin_JPCM2018,Mokshin_JPCM2018/Corrigendum}.  A similar consistent, albeit more simplified, description until now could only be realized in the framework of the viscoelasticity model~\cite{Copley/Lovesey}.

The theory yields the proper asymptotic behavior in both the short-wave limit, corresponding to free particle motion, and the long-wave hydrodynamic limit and allows calculating hydrodynamic characteristics. The key condition determing the correspondence between time scales of relaxation processes in liquids allows acounting for the presence of high-speed processes, which are usually ignored in hydrodynamic theory and some of its generalizations, in the framework of the presented theoretical description. The theory satisfies a complete set of so-called sum rules of the spectral density $C_T(k,\omega)$. It generalizes the known viscoelasticity model~\cite{Copley/Lovesey}, whose key parameters can be calculated in the framework of the presented theory (also see Sec. 3.1. in~\cite{Mokshin_JPCM2018}).

In this theory, we did not use any assumptions that some particular ``relaxation'' and ``oscillation'' regimes (or modes) form any features of the transverse collective particle dynamics, in turn determining some regions of the spectral density $C_T(k,\omega)$ of the transverse particle current. Therefore, it is totally unnecessary to separate or isolate any components of this spectral density for interpreting an experimentally measured or a numerically modeled spectral density.


\begin{thebibliography}{99}
		
\bibitem{Frenkel_book} J. Frenkel, \textit{Kinetic Theory of Liquids}, Clarendon Press, Oxford (1946)
		
\bibitem{Bol_Phys_Enciclopedy} A. M. Prokhorov et al., eds., \textit{Encyclopedic Dictionary of Physics} [in Russian], Sovet. Entsiklopediya, Moscow (1983).
		
\bibitem{Phys_Enc} A. M. Prokhorov et al., eds., \textit{Encyclopedia of Physics} [in Russian], Vol. 2, Bol'shaya Rossiiskaya Entsiklopediya,
		Moscow (1998).
		
\bibitem{Trachenko2016} K. Trachenko and V.V. Brazhkin, \textit{Rep. Prog. Phys.}, \textbf{79} (2016), 016502, 36 pp.
		
\bibitem{Tareyeva2018} E. E. Tareyeva, Yu. D. Fomin, E. N. Tsyok, and V. N. Ryzhov, \textit{Theor. Math. Phys.}, \textbf{194}:1 (2018), 148-156.
		
\bibitem{hypersound_experiments} A.V. Granato, \textit{J. Phys. IV France}, \textbf{06} (1996), C8, 9 pp.
		
\bibitem{Levesque/Verlet_1973} D. Levesque, L. Verlet, J. K\"{u}rkijarvi, \textit{Phys. Rev. A}, \textbf{7}:5 (1973), 1690-1700.
		
\bibitem{Sjogren1978} L. Sj\"{o}gren, \textit{Ann. Phys.}, \textbf{110}:1 (1978), 173-179.
		
\bibitem{Donko} Z. Donk\'{o}, G.J. Kalman, P. Hartmann, \textit{J. Phys.: Condens. Matter}, \textbf{20}:41 (2008), 413101, 35 pp.
		
\bibitem{Khrapak2019} S.A. Khrapak, A.G. Khrapak, N.P. Kryuchkov, S.O. Yurchenko, \textit{J. Chem. Phys.}, \textbf{150}:10 (2019), 104503, 8 pp.
		
\bibitem{Ryltsev2013} R.E. Ryltsev, N.M. Chtchelkatchev, V.N. Ryzhov, \textit{Phys. Rev. Lett.}, \textbf{110}:2 (2013), 025701, 5 pp.
		
\bibitem{Gonzalez2017} B.G. del Rio, L.E. Gonz\'{a}lez, \textit{Phys. Rev. B}, \textbf{95}:22 (2017), 224201, 15 pp.
		
\bibitem{Jakse/Bryk2019} No\"{e}l Jakse, T. Bryk, \textit{J. Chem. Phys.}, \textbf{151}:3 (2019), 034506, 8 pp.
		
\bibitem{Jones2016}  M. Ropo, J. Akola, R.O. Jones, \textit{J. Chem. Phys.}, \textbf{145}:18 (2016), 184502, 8 pp.
		
\bibitem{Fomin2019} Y.D. Fomin, E.N. Tsiok, V.N. Ryzhov, V.V. Brazhkin, \textit{J. Molec. Liq.}, \textbf{287} (2019), 110992, 4 pp.
		
\bibitem{Wang2019} L. Wang, C. Yang,  M.T. Dove,  A.V. Mokshin, V.V. Brazhkin, K. Trachenko,  \textit{Sci. Rep.}, \textbf{9} (2019), 755, 9 pp.
		
\bibitem{Hosokawa_2015} S. Hosokawa, M. Inui, Y. Kajihara, S. Tsutsui, A.Q.R. Baron,  \textit{J. Phys.: Condens. Matter}, \textbf{27}:19 (2015), 194104, 7 pp.
		
\bibitem{Egelstaff_book} P.A. Egelstaff, \textit{An Introduction to the Liquid State}, Academic Press, New York, 1967.
		
\bibitem{Burkel_review} E. Burkel, H. Sinn, \textit{J. Phys.: Condens. Matter}, \textbf{6} (1994), A225-A225.
		
\bibitem{Hosokawa_Ga} S. Hosokawa, M. Inui, Y. Kajihara, K. Matsuda, T. Ichitsubo, W.C. Pilgrim, H. Sinn, L.E. Gonz$\acute{a}$lez, D.J. Gonz$\acute{a}$lez, S. Tsutsui, A.Q.R. Baron, \textit{Phys. Rev. Lett.}, \textbf{102}:10 (2009), 105502, 4 pp.
		
\bibitem{Hosokawa2013} S. Hosokawa, S. Munejiri, M. Inui, Y. Kajihara, W.-C. Pilgrim, Y. Ohmasa, S. Tsutsui, A.Q.R. Baron, F. Shimojo and K. Hoshino,
\textit{J. Phys. Condens. Matter}, \textbf{25}:11 (2013) 112101, 5 pp.
		
\bibitem{Monaco_PNAS} V.M. Giordano, G. Monaco, \textit{Proc. Natl. Acad. Sci. USA}, \textbf{107}:51 (2010), 21985-21989.
		
\bibitem{Monaco_PRB} V.M. Giordano, G. Monaco, \textit{Phys. Rev. B}, \textbf{84}:5 (2011), 052201, 4 pp.
		
%% Viscoelastic model for transverse dynamics
\bibitem{MacPhail/Kivelson} R.A. MacPhail, D. Kivelson, \textit{J. Chem. Phys.}, \textbf{80}:5 (1984), 2102-2114.
		
\bibitem{Bryk_PRE2000} T. Bryk, I. Mryglod, \textit{Phys. Rev. E}, \textbf{62}:2 (2000), 2188-2199.
		
%% Generalized collective modes approach
\bibitem{Mryglod_CMP1994} I.P. Omelyan, I.M. Mryglod, \textit{Condens. Matter Phys.}, \textbf{N4} (1994), 128-160.
		
%% Maxwell-Frenkel approach
\bibitem{Trachenko2017} K. Trachenko, \textit{Phys. Rev. E}, \textbf{96}:6 (2017), 062134, 5 pp.
\bibitem{Baggioli2020} M. Baggioli, M. Vasin, V. Brazhkin and K. Trachenko, \textit{Phys. Rep.}, \textbf{865} (2020), 1-44.
		
%% Separate mode analysis versus the two-oscillator model.
\bibitem{Yurchenko2019} N.P. Kryuchkov, L.A. Mistryukova, V.V. Brazhkin, S.O. Yurchenko, \textit{Sci. Rep.}, \textbf{9} (2019), 10483, 12 pp.
		
\bibitem{Yurchenko_SC_2019} N.P. Kryuchkov, V.V. Brazhkin, S.O. Yurchenko, \textit{J. Phys. Chem. Lett.}, \textbf{10} (2019),  4470-4475.
		
\bibitem{Yurchenko_1} E.V. Yakovlev, N.P. Kryuchkov, P.V. Ovcharov, A.V. Sapelkin, V.V. Brazhkin, S.O. Yurchenko, \textit{J. Phys. Chem. Lett.}, \textbf{11} (2020), 1370-1376.
		
\bibitem{Fomin_JPCM_2016} Yu.D. Fomin, V.N. Ryzhov, E.N. Tsiok, V.V. Brazhkin, K. Trachenko, \textit{J. Phys.: Condens. Matter}, \textbf{29} (2017), 059501, 1 pp.
		
\bibitem{Brazhkin_2018} V.V. Brazhkin, Yu.D. Fomin, V.N. Ryzhov, E.N. Tsiok, K. Trachenko, \textit{Physica A: Statistical Mechanics and its Applications}, \textbf{509} (2018), 690-702.
		
\bibitem{Mokshin_PRE2001} R.M. Yulmetyev, A.V. Mokshin, P. H\"{a}nggi, V.Yu. Shurygin, \textit{Phys. Rev. E}, \textbf{64}:5 (2001), 057101, 4 pp.
		
\bibitem{Mokshin2002} R.M. Yulmetyev, A.V. Mokshin, P. H\"{a}nggi, V.Yu. Shurygin, \textit{JETP Lett.}, \textbf{76}:3 (2002), 147-150.
		
\bibitem{Mokshin_JPCM2018} A.V. Mokshin, B.N. Galimzyanov, \textit{J. Phys.: Condens. Matter}, \textbf{30} (2018), 085102, 17 pp.
		
\bibitem{Mokshin_JPCM2003} R.M. Yulmetyev, A.V. Mokshin, T. Scopigno, P. H\"{a}nggi, \textit{J. Phys.: Codens. Matter}, \textbf{15} (2003), 2235-2257.
		
\bibitem{Mokshin_JCP2004} A.V. Mokshin, R.M. Yulmetyev, P. H\"anggi, \textit{J. Chem. Phys.}, \textbf{121}:15 (2004), 7341-7346.
		
\bibitem{Mokshin_JETP2006} A. V. Mokshin, R. M. Yulmetyev, R. M. Khusnutdinov, and P. H\"{a}nggi, \textit{JETP}, \textbf{103}, 841-849, (2006).
		
\bibitem{Mokshin_JPCM2007} A.V. Mokshin, R.M. Yulmetyev, R.M. Khusnutdinoff, P. H\"{a}nggi, \textit{J. Phys.: Condens. Matter}, \textbf{19} (2007), 046209,
		16 pp.
		
\bibitem{Khusnutdinoff2020} R.M. Khusnutdinoff, C. Cockrell, O.A. Dicks, A.C.S. Jensen, M.D. Le, L. Wang, M.T. Dove, A.V. Mokshin, V.V. Brazhkin, and K. Trachenko, \textit{Phys. Rev. B}, \textbf{101}:21 (2020), 214312, 9 pp.
		
\bibitem{Ryzhov_UFN2008} V. N. Ryzhov, A. F. Barabanov, M. V. Magnitskaya, and E. E. Tareeva, \textit{Phys. Usp.}, \textbf{51}, 1077-1083 (2008).
		
\bibitem{Hansen/McDonald} J.P. Hansen, I.R. McDonald, \textit{Theory of Simple Liquids} Academic Press, London, 2006.
		
\bibitem{Zwanzig2001} R. Zwanzig, \textit{Nonequilibrium statistical mechanics}, Clarendon Press, Oxford, 2001.
		
\bibitem{Mokshin/Yulmetyev} A. V. Mokshin and R. M. Yulmetyev, \textit{Microscopic Dynamics of Simple Liquids} [in Russian], Tsentr Innovatsionnykh Tekhnologii, Kazan (2006).
		
\bibitem{Klumov} B. A. Klumov, \textit{Phys. Usp.}, \textbf{53}, 1053-1065 (2010).
		
\bibitem{Balucani2003} U. Balucani, M.H. Lee, V. Tognetti, \textit{Phys. Rep.}, \textbf{373}:6 (2003), 409-492.
		
\bibitem{Reed_Book} M. Reed and B. Simon, \textit{Methods of Modern Mathematical Physics}, Vol. 1: \textit{Functional Analysis}, Acad. Press, New York (1972).
		
\bibitem{Plakida2005} A. A. Vladimirov, D. Ihle, and N. M. Plakida, \textit{Theor. Math. Phys.}, \textbf{145}, 1576-1589 (2005).
		
\bibitem{Lee2000} M.H. Lee, \textit{Phys. Rev. E}, \textbf{62}:2 (2000), 1769-1772.
		
\bibitem{Mokshin2005} A.V. Mokshin, R.M. Yulmetyev, P. H\"{a}nggi, \textit{Phys. Rev. Lett.}, \textbf{95}:20 (2005), 200601, 4 pp.
		
\bibitem{Mokshin2015} A. V. Mokshin, \textit{Theor. Math. Phys.}, \textbf{183}, 449-477 (2015).
		
\bibitem{Bogolyubov_book} N. N. Bogolyubov, \textit{Problems of Dynamical Theory in Statistical Physics} [in Russian], Gostekhizdat, Moscow (1946); English transl. (Stud. Statist. Mech., Vol. 1), North-Holland, Amsterdam (1962).
		
\bibitem{Gotze2009} W. G\"{o}tze, \textit{Complex Dynamics of Glass-Forming liquids: A Mode-Coupling Theory}, Oxford Univ. Press, Oxford, 2009.
		
\bibitem{Resibua/Lener} P. Resibua and M. De Lener, \textit{Classical Kinetic Theory of Liquids and Gases} [in Russian], Moscow, Mir, (1980).
		
\bibitem{Mountain_1966} R. Mountain, \textit{Rev. Mod. Phys.}, \textbf{38}:1 (1966), 205-214.
		
\bibitem{Scopigno2000} T. Scopigno, U. Balucani, G. Ruocco and F. Sette, \textit{J. Phys.: Condens. Matter}, \textbf{12} (2000), 8009-8034.
		
\bibitem{Murtazayev1999} I. K. Kamilov, A. K. Murtazaev, and Kh. K. Aliev, \textit{Phys. Usp.}, \textbf{42}, 689-709 (1999)
		
\bibitem{Khusnutdinoff2018} R. M. Khusnutdinoff, B. N. Galimzyanov, and A. V. Mokshin, \textit{JETP}, \textbf{126}, 83-89 (2018).
		
\bibitem{Gonzalez1993} L.E. Gonz$\acute{a}$lez, D.J. Gonz$\acute{a}$lez, M. Silbert and J.A. Alonso, \textit{J. Phys.: Condens. Matter}, \textbf{5} (1993), 4283-4298.
		
\bibitem{Mokshin2012} A. V. Mokshin, A. V. Chvanova, and R. M. Khusnutdinov, \textit{Theor. Math. Phys}, \textbf{171}, 541-552 (2012).
		
\bibitem{Waseda} Y. Waseda, \textit{The Structure of Non-Crystalline Materials: Liquids and Amorphous Solids}, McGraw-Hill, New York, 1980.
		
\bibitem{link_1} R.W. Ohse (Ed.), \textit{Handbook of Thermodynamic and Transport Properties of Alkali Metals}, Intern. Union of Pure and Applied
		Chemistry Chemical Data Series No. 30. Oxford: Blackwell Scientific	Publ., 1985, pp. 987.
		
\bibitem{Mokshin_JPCM2018/Corrigendum} A.V. Mokshin, B.N. Galimzyanov, \textit{J. Phys.: Condens. Matter}, \textbf{31}:20 (2019), 209501, 1 pp.
		
\bibitem{Copley/Lovesey} J.R.D. Copley, S.W. Lovesey, \textit{Rep. Prog. Phys.}, \textbf{38} (1975), 461-563.
		
\end{thebibliography}
\end{document}